# Enhancing Image Resolution: A Simulation Study and Sensitivity Analysis of System Parameters for Resourcesat-3S/3SA

Ankur Garg [†], Meenakshi Sarkar, S. M. Moorthi, Debajyoti Dhar
*Optical Data Processing Group, Signal and Image Processing Area,*
*Space Applications Centre, Ahmedabad*

October 22, 2024

## 1  Introduction

Resourcesat-3S/3SA, a forthcoming Indian satellite, will house both Aft and Fore payloads, offering look angles of -5° and 26° respectively to capture stereo images. Operating at a proposed altitude of 632.6 km, the panchromatic (PAN) band is slated to provide an impressive Ground Sampling Distance (GSD) of 1.25 m, covering a swath of approximately 60 km. Given the constraints of limited space and the need for a substantial swath and high resolution, achieving these specifications simultaneously in hardware is deemed unfeasible. Thus, a compromise is reached by maintaining an Instantaneous Geometric Field of View (IGFOV) of 2.5 m, while preserving the GSD at 1.25 m both along and across track.

Achieving along track sampling necessitates precise timing corresponding to 1.25 m intervals, while for across track sampling, two staggered pixel arrays are employed. To bolster the Signal-to-Noise Ratio (SNR), Time Delay and Integration (TDI) process has been proposed. This involves the incorporation of two five-stage TDI arrays, referred to as subarrays, situated approximately 80 $\mu$m (equivalent to 10 scan lines) apart along the track. Additionally, Sub-Array 2 will be 4 $\mu$m (or 0.5 pixels) staggered relative to Sub-Array 1 in the across-track direction, thereby achieving 1.25m sampling in the across track direction.

The technique of super resolution (SR) is employed to generate a high-resolution composite image from multiple low-resolution captures of the same scene. SR leverages sub-pixel shifts in these images to amalgamate information and enhance resolution. This is particularly effective when the input low-resolution images are under-sampled and carry high-resolution details in an aliased form. By utilizing phase information from an adequate number of observations, it is possible to disentangle the aliased high-frequency data from the true low-frequency content, enabling the reconstruction of a full-resolution image. This methodology has been successfully implemented in satellite remote sensing, as demonstrated by missions such as SPOT-5 (HRG), SkySat and DubaiSat-1, resulting in significantly improved image resolutions. The present study delves into the various factors influencing the ultimate resolution achievable in this RS3S/3SA configuration. Through a Monte Carlo simulation approach, a sensitivity analysis is conducted to discern the impact of these factors. Importantly, the simulation methodology outlined in this report holds broader applicability and can be adapted for similar missions in satellite remote sensing.

---

[†]Corresponding author : agarg@sac.isro.gov.in

# 2 Overview

## 2.1 Super-Resolution

Achieving the highest possible image resolution without distortion is always a primary goal in acquisition systems. However, attaining high resolution often necessitates a sensor with a large aperture and large focal length. An alternative approach to enhance image quality is through post-processing on the ground. This is where super-resolution techniques come into play, enabling the generation of a high-resolution signal from multiple low-resolution inputs.

A fundamental requirement for any super-resolution approach to be effective, from a system design perspective, is that the input low-resolution images should be significantly under-sampled, thus containing aliasing. This necessitates a high telescope modulation transfer function at the detector Nyquist frequency. This aliasing results in the high-frequency content of the desired reconstruction image being embedded in the low-frequency content of each observed image. Another crucial requirement is that the low-resolution images should exhibit variation in their phase, i.e., they should be shifted by sub-pixel amounts. This phase information can then be utilized to distinguish the aliased high-frequency content from the true low-frequency content, enabling the accurate reconstruction of the full-resolution image.

Over the past few decades, a multitude of super-resolution methods have been documented in the literature. These methods can be broadly categorized into various groups, including Fourier-based approaches[1], non-uniform interpolation-based techniques[2], iterative back projection-based methods[3], Projection on Convex Sets (POCS)[4], Maximum a Posteriori (MAP) estimators[5], and more recently, learning-based methods[6].

In the realm of image enhancement, MAP-based methodologies occupy a distinctive position, distinguished by their capacity to integrate prior knowledge through regularization techniques. These approaches are typically comprised of two integral components. The first element centers on the fidelity term, where the utilization of the L2 norm, rooted in Gaussian noise assumptions, is customary. Yet, it's important to note that these methods can be vulnerable to outliers like salt and pepper noise or transmission loss, as they disproportionately penalize errors in a quadratic fashion. In scenarios featuring outliers, the application of L1 norms based on assumptions of white Laplacian noise has exhibited superior performance[7], offering heightened resilience against registration inaccuracies. Nevertheless, it's worth considering that while L1 norms can be effective, they may introduce more observation error, lack differentiability, and potentially lead to an excessive smoothing of images. This could potentially result in a loss of vital information, particularly in the context of remote sensing imagery. Recently, M-estimators like Huber functions[8] have emerged as versatile alternatives, capable of acting as both L1 and L2 norms based on a threshold parameter. The second component of the MAP estimate involves the incorporation of a prior model based on assumptions about the desired solution. Regularizers like L1-based Tikhonov (TV)[9] norms and sparsity-based L1 norms, including Bilateral TV (BTV)[7], have been proposed. Adaptive versions of these norms, designed to reduce noise while enhancing edge information, have also been introduced, such as those based on difference curvature for BTV[11] and TV[12].

Deep learning-based approaches have surged in popularity in recent years in image super resolution task. However, it's important to acknowledge that their applicability is specialized to the specific types of images they are trained on. Another challenge arises in generating high-resolution and low-resolution images for training. Since many of these methods rely on simulating expected data from existing sensor data, their performance on real-world datasets can vary significantly. These methods may produce outputs that are plausible but not entirely accurate representations of real data.

Some research endeavors have aimed to train deep learning models using simulated datasets, where Low-Resolution (LR) images are generated using a predefined degradation model [13]. However, the effectiveness of these models may decline considerably if the real low-resolution input deviates from the simulated degradation model. Other approaches incorporate real HR images obtained from different satellites to directly guide the Super-Resolution (SR) process of a coarse resolution satellite like Sentinel-2 [[14], [16], [15]]. Nevertheless, acquiring HR ground truth images can be financially demanding. Furthermore, the utilization of HR images from diverse satellites introduces challenges, including spectral response variations, differences in acquisition viewpoint and time. These complexities complicate the dataset creation process and have a detrimental impact on overall performance. Also they suffer from "hallucination" effect where the model may generate information which may not be present in the real world. Certain methodologies, such as the one outlined in [17], employ self-supervision for model training but exhibit a pronounced reliance on the specific hardware imaging acquisition strategies. As an illustration, the utilization of overlapping regions is restricted to early products within the Sentinel-2 processing pipeline, specifically in the level-1B (L1B) products. These early products introduce a notable interband parallax, a consequence of the hardware design of detectors aimed at enhancing the resolution of Sentinel-2 images.

In operational satellite data processing, super resolution has been used in SPOT-5 HRG[18] to increase the spatial resolution from 5 meters(m) to 3m. Skybox[19] also uses algorithm to generate high resolution channel from video acquisitions. Both the satellites acquire images from lower earth orbit. These super-resolution algorithm works on the assumption that the low-resolution images are single channel and come from same radiometry.

Given these constraints, the performance of deep learning-based methods is not always guaranteed, which makes them less suitable as benchmark algorithms for conducting comprehensive system-level studies. In contrast, MAP-based methods are firmly grounded in well-established physical imaging principles. This lends them a higher degree of reliability and interpretability. For these reasons, MAP-based methods have been selected as the benchmark algorithms for the analysis presented in this report.

## 2.2 Observation Model

This reports adopts the L2 norm with a bilateral filter as the baseline algorithm for simulation. The observation model used to derive the low-resolution PAN image from the high-resolution PAN band is expressed in Equation 1:

$$Y_k = DM_kB_kX + N_k \tag{1}$$

Here, subscript $k$ denotes the $k$-th MX band, $Y_k$ represents the low-resolution MX band of size $N \times M$, $X$ is the high-resolution PAN band of size $sN \times sM$, with $s$ being the scaling factor. $B_k$ denotes the high-resolution point spread function, representing the blur the high-resolution image undergoes before sampling to produce this band. $M_k$ is the motion matrix, containing misregistration information between different bands. $D$ is the decimation matrix, and $N_k$ is the noise matrix corresponding to the low-resolution PAN image.

Given $n$ such low-resolution images, the high-resolution PAN band $X$ can be derived by minimizing Equation 2:

$$\arg\min \sum_{k=1}^{n} ||Y_k - DM_kB_kX||_p^p + \phi(X) \tag{2}$$

In this equation, the first term represents the likelihood term with norm $p$, and $\phi(X)$ represents the prior probability of the desired solution in the MAP framework. $p = 2$ corresponds to the Gaussian noise model assumption, $p = 1$ corresponds to the Laplacian noise model assumption, and $p < 1$ corresponds to hyper-Laplacian assumptions.

As discussed previously, the prior model $\phi(X)$ can take the form of $L2$ norm, $L1$ norm, BTV norm, or $Lp$ norm where $p < 1$ [22] to enhance sparsity. The complete minimization function for the super-resolution framework is described in Equation 3:

$$\begin{aligned}\hat{X} = \arg\min \sum_{k=1}^{n} ||Y_k - DM_kB_kX||_2^2 + \\ \lambda w \sum_{l=-P}^{P} \sum_{m=0, l+m>0}^{P} \alpha^{|m|+|l|} ||X - S_h^l S_v^m X||\end{aligned} \tag{3}$$

Here, $\lambda$ is the parameter controlling the strength of the regularization parameter compared to the likelihood term.

To solve the above equation, an iterative steepest gradient algorithm is employed. The iterative solution is derived by differentiating the equation with respect to $X$ and setting the derivative to zero, resulting in the following iterative solution:

$$\hat{X}_{n+1} = \hat{X} - \beta\Big\{\sum_{k=1}^{n} M_k^T B_k^T D_k^T (Y_k - DM_k B_k X) + \\ \lambda w \sum_{l=-P}^{P} \sum_{m=0, l+m>0}^{P} \alpha^{|m|+|l|}(I - S_h^{-l} S_v^{-m})\text{sign}(\hat{X} - S_h^l S_v^m \hat{X})\Big\} \quad (4)$$

In this equation, $\beta$ represents the learning rate, which is adaptively adjusted for faster convergence.

## 2.3 Resolution-Measurement

Image resolution refers to the capability of an imaging system to distinguish fine details within an image. It is determined by the combined performance of the optics and the detector of the camera. Generally, as the spatial frequencies in a scene increase, the contrast in the camera's output decreases. This relationship is defined by the modulation transfer function (MTF). At a certain spatial frequency, the contrast diminishes to a point where noise-induced modulation surpasses signal-induced modulation. This marks the maximum cutoff frequency, beyond which scene details become indistinguishable from noise.

ISO 12233 provides standardized methods for automatically measuring contrast, and one robust approach employs Siemens's star target. This method has been employed for resolution measurement in our study since it allows reliably assessing sensor resolution due to its broad range of spatial frequencies. The modulation for each spatial frequency can be measured from the target by calculating the amplitude at varying spatial frequencies. Figure 1 displays an image of a Siemens's star target, which comprises a total of 144 cycles. It can be subdivided into eight segments for orientation-specific image processing.

Supposing $\alpha$ is the angle where a particular pixel is present with respect to the center of the image. The angle $\alpha$ is given by:

$$\alpha = tan^{-1}\frac{x}{y} \quad (5)$$

where $x$, $y$ represent the position of the pixel and $x = 0$ and $y = 0$ represents the center of the star. The intensity at the angle $\alpha$ is given by

$$I(\alpha) = a + \beta.cos(\frac{2\pi}{g}.(\alpha - \alpha_0)) \quad (6)$$

where $g = \frac{2\pi r_{pixel}}{N_y}$ is the cycle length in pixels. $\alpha$ is the mean intensity value and $\beta$ is the amplitude of intensity modulations. The modulation at a particular frequency is given by:

$$M = \frac{I_{max} - I_{min}}{I_{max} + I_{min}} = \frac{\beta}{\alpha} \quad (7)$$

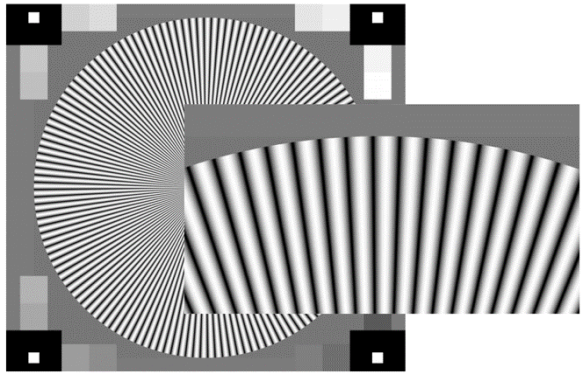

Figure 1: Siemen's Spoke Target

## 2.4 Noise-Equivalent Modulation

Let $s$ be the signal level and $n =$ be the noise standard deviation at this signal level, then the noise equivalent modulation(NEM) is given by:

$$NEM = \frac{(s+2*n) - (s-2*n)}{s} = \frac{4*n}{s} \tag{8}$$

The spatial frequency where modulation transfer function meets the noise equivalent modulation is the achieved resolution. The signal level is assumed to be equivalent to 15% albedo. The noise is assumed to be white (same to all frequencies). Figure 2 provides an illustrative example where the derived MTF from a star target intersects with the NEM. In this scenario, the signal level is set at 300 counts, and the noise has a standard deviation of 5, resulting in an NEM equivalent to 0.66%. The frequency at which the MTF intersects with the NEM is 0.43 cycles per pixel, yielding an achieved resolution of 1.48 meters. This is because the Nyquist frequency of 0.5 cycles per pixel corresponds to a resolution of 1.25 meters.

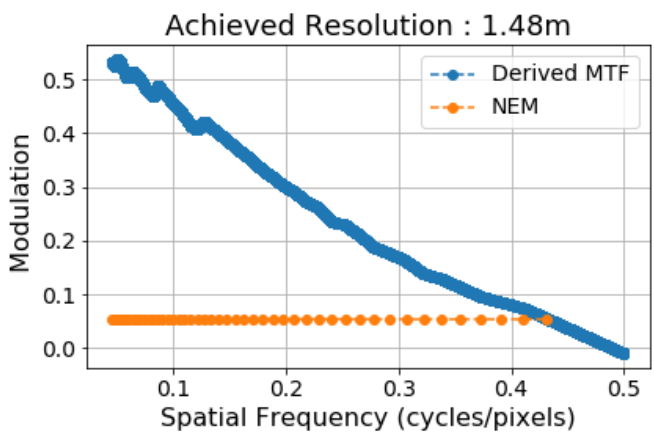

Figure 2: Noise Equivalent Modulation

## 2.5 Raw Image Simulator:

The Raw Image Simulator plays a pivotal role in this Monte Carlo experiment. It operates as a model that takes as input a high-resolution spoke target image, coupled with an array of imaging system parameters. Subsequently, it employs these parameters to generate two low-resolution subarray images, incorporating all degradation factors specified by the system parameters. These resultant low-resolution images then serve as the foundational input for the subsequent super resolution process. In essence, the Raw Image Simulator serves as a crucial intermediary step in the experimental setup, accurately emulating the imaging conditions and enabling the evaluation of super resolution performance. The flowchart of the simulator is shown in Figure 3.

# 3 Factors Effecting Super Resolution Performance

The generation of low-resolution images from the payload is influenced by numerous factors. These factors have a direct impact on the quality and fidelity of the final high-resolution image produced. The various parameters that play a crucial role in influencing the performance of super resolution are outlined below:

## 3.1 Optical Modulation Transfer Function (MTF)

An optical system with a fixed-size aperture can never form a point source due to the wave nature of light. Even when the aberrations are absent, there is a minimum blur diameter formed. The best

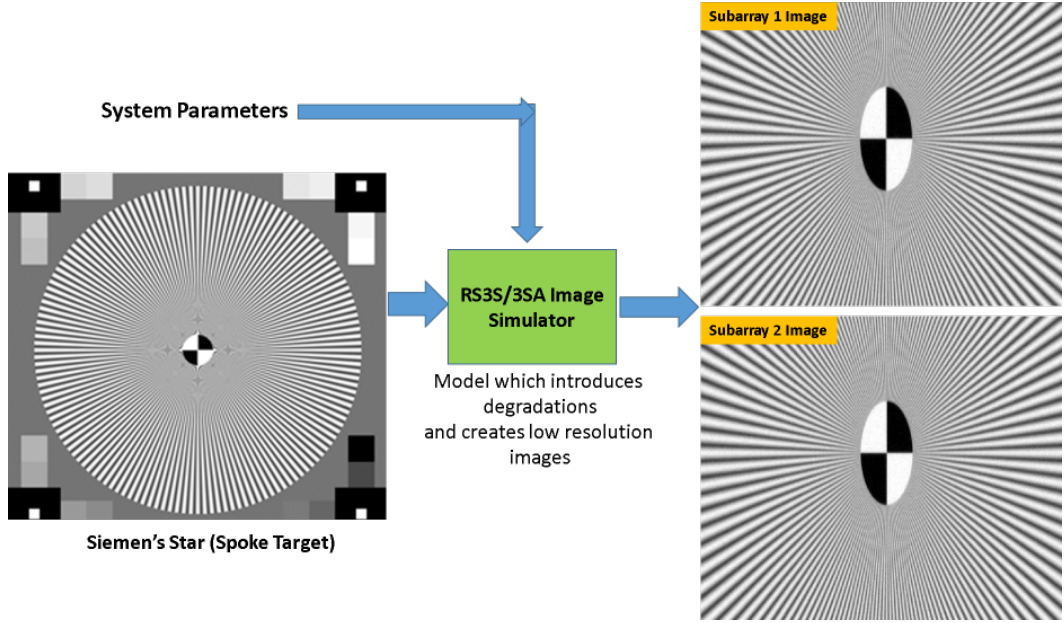

Figure 3: RS3S/3SA Raw Image Simulator

performance that the system is capable of is thus called diffraction limited [21]. The diameter of the diffraction blur is

$$d_{diffraction} = 2.44\lambda F_{num} \quad (9)$$

where the $F_{num}$ is given by $f/D$, where $f$ is called the focal length and $D$ is the aperture size. The optical system has the cutoff frequency, $\sigma_{cutoff} = 1/(\lambda_{num})$. For the case of a circular aperture of diameter D, but the MTF has a functional form:

$$MTF(\Sigma) = \frac{2}{\pi}cos^{-1}\Sigma - \Sigma[1 - \Sigma^2]^{1/2} \quad (10)$$

where $\Sigma = (\sigma/\sigma_{cutoff})$ for $\sigma < \sigma_{cutoff}$ and

$$MTF(\Sigma) = 0 \quad (11)$$

for $\sigma > \sigma_{cutoff}$

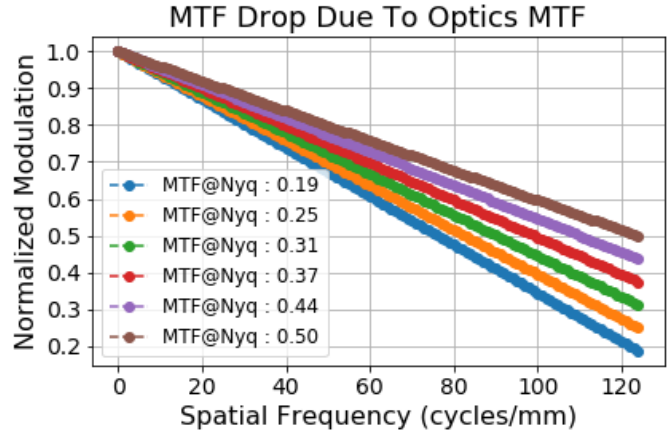

Figure 4: Optical Modulation Transfer Function

The realizable optical MTF typically falls short of the diffraction-limited MTF due to various factors such as optical aberrations, fabrication imperfections, and alignment-related issues. Specifically, the anticipated value of the optics MTF in RS3S/3SA is 58% at a spatial frequency of 62.5 lp/mm, and 30% at 125 lp/mm.

For simulation purposes, a range of optical MTF values have been taken into account. Figure 4 illustrates the profiles of these realizable optical modulation functions for the RS3S/3SA system. This provides a visual representation of how the MTF varies across different spatial frequencies, and how it impacts the system's imaging capabilities.

## 3.2 Detector Footprint MTF and Detector Sampling MTF

A square detector of size $wxw$ performs spatial averaging of the scene irradiance that falls on it. When we analyze the situation in one dimension, we find the integration of the scene irradiance f(x) over the detector surface is equivalent to a convolution of f(x) and the rect function that describes the detector responsivity[21].

$$g(x) = \int_{-w/2}^{w/2} f(x) * rect(x/w) \tag{12}$$

By the convolution theorun, the above equation is equivalent to filetering in the frequency domain by a transfer function

$$MTF_{footprint}(\sigma) = |sinc(\sigma w)| = |\frac{sin(\pi \sigma w)}{(\pi \sigma w)}| \tag{13}$$

This shows that smaller the sensor dimension is, the broader the transfer function.
Sampling impulse response is a rectangle function whose widths are equal to the sampling intervals in each direction.

$$h_{sampling}(x, y) = rect(x/x_{samp}, y/y_{samp}) \tag{14}$$

The aforementioned equation highlights that when sampling is more widely spaced, it leads to images with lower quality. Figure 5 provides a visual representation of the Modulation Transfer Function (MTF) profile for both the detector aperture and the sampling process specific to the RS3S /3SA system. This profile showcases how the MTF varies with different spatial frequencies, offering insights into the system's imaging characteristics.

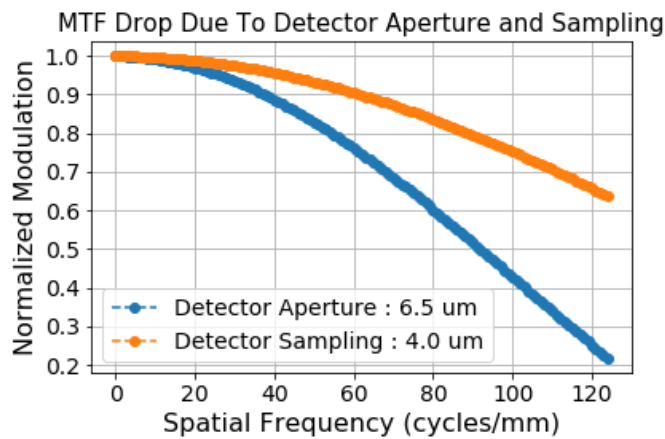

Figure 5: Detector Aperture and Sampling

## 3.3 Signal To Noise Ratio (SNR)

In a remote sensing system, noise arises from various sources: Photon or Poisson Noise occurs due to the discrete nature of light, resulting in random fluctuations in photon arrivals at the sensor. This is particularly noticeable in low-light conditions. Thermal Noise, also known as thermal electronic noise, originates from the random movement of electrons within the electronic components of the sensor. It is influenced by the system's temperature. Read Noise is introduced during the process of converting the analog sensor signal to a digital form. It is a fixed component associated with the electronic readout process. Fixed Pattern Noise emerges from variations in the behavior of individual pixels or groups of pixels on the sensor. These variations can be due to manufacturing imperfections or intrinsic characteristics of the detector.

The Signal-to-Noise Ratio (SNR) is a critical metric used in imaging to gauge the system's sensitivity. It is calculated as the ratio of the mean signal level to the variance of the noise at that signal level. A higher SNR signifies a system capable of producing clearer and more detailed images. SNR directly influences the Noise Equivalent Modulation (NEM) of the system. NEM represents the modulation level at which the signal becomes indistinguishable from the noise. A higher SNR leads to a lower NEM, which, in turn, enhances the system's resolution capabilities. Conversely, a lower SNR can limit the system's ability to accurately resolve fine details. In summary, a higher SNR is preferred for obtaining high-quality images with improved resolution, as it effectively mitigates the influence of noise in the acquired data. In RS3S/3SA, the expected value of SNR at 15% albedo which corresponds to 300 counts is 60.

## 3.4 Clock Smear / Motion Blur

One of the unique aspects of the TDI mode of operation is that the signal charge in the imaging array is moved in discrete steps, whereas the mechanical scanning can be almost linear or proceed in discrete steps, depending on the method used for achieving mechanical displacement. This results in a mode in which the signal charge is always either lagging behind the image or is ahead of the image by a sub pixel amount and is never in total synchrony with the optical image-namely, in a mode in which the signal charge moves in a "jerky" motion, while the scanned optical image moves almost linearly or in discrete steps with step sizes dierent from those of the charge packet motion. This discrete motion of the charge causes misregistration between the image and the imaging potential wells, and results in a degradation of the MTF in the mechanical scanning (along-track) TDI direction. In order to calculate the degradation in the MTF due to discrete charge motion, we consider (a) a pattern of charge movement which is dependent on the clocking method for the imaging parallel array and, (b) mechanical scanning motion[20]. For each charge transfer, the image will move a distance of $(p/n\phi)$, where p is the center-to-center CCD stage spacing, and $n\phi$ is the number of moves per stage. The MTF is given by

$$MTF_{discrete} = sinc(\frac{\pi}{2} \frac{f}{f_N} \frac{1}{n\phi}) \tag{15}$$

Essentially, the greater the number of moves per stage, the smoother the motion of the signal charge, and thus the less degradation in the MTF because of discrete charge motion. Figure 6 shows the MTF drop in RS3S/3SA for various values of clock phases. The expected value of motion MTF with 1 clock phase at 62.5lp/mm is 90% and 125 lp/mm is 64%.

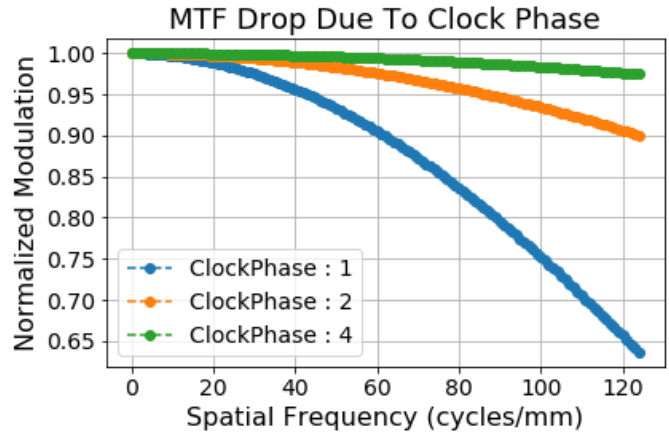

Figure 6: MTF Drop due to Smear/ Clock Phase

## 3.5 Satellite Jitter

Satellite Jitter is the variation of imaging line of sight within the integration time. These high frequencies can contribute to image degradation, particularly in terms of a reduction in the Modulation Transfer Function (MTF). Figure 7 illustrates the decline in MTF with varying magnitude of jitter. This drop in MTF indicates the impact of jitter on the system's ability to faithfully reproduce high-frequency details. Jitter can be effectively modeled using a Gaussian function with a specified amplitude. In the case of RS3S/3SA, the anticipated value of jitter is 0.1 pixels. This level of jitter is expected to result in a 95% reduction in MTF at a spatial frequency of 125 lp/mm. This demonstrates the significant influence of jitter on the imaging performance of the RS3S system.

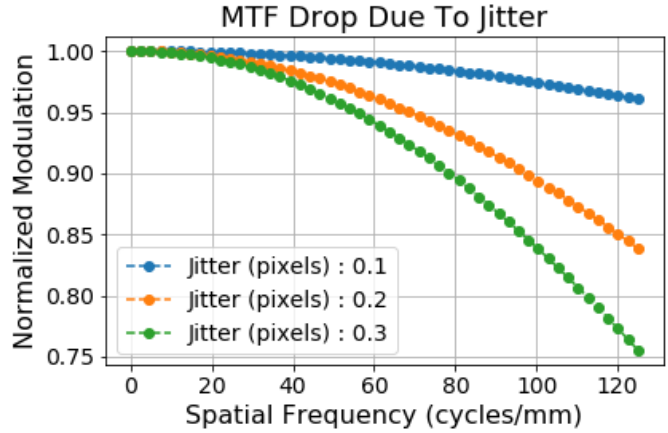

Figure 7: MTF Drop due to Jitter

## 3.6 Misalignment Between Subarrays

The misalignment between sub-arrays has a direct impact on the phase shift between the low-resolution images. In an ideal scenario with only two low-resolution samples, they should ideally be spaced 0.5 pixels apart. This specific spacing is crucial for super resolution performance because it allows for the extraction of cross-phase information from the low-resolution images. When the actual positioning deviates from this ideal scenario, it results in a degradation of super resolution performance. This is primarily because the necessary cross-phase information cannot be accurately derived from the low-resolution images. In essence, misalignment disrupts the crucial relationship between the samples, making it more challenging to reconstruct a high-resolution image with precision. In RS3S/3SA, the intended across-track shift is achieved by strategically placing two staggered detectors at the focal plane with a separation of 0.5 pixels. However, it's worth noting that due to factors like differences in pitch angle and potential platform drift, this value may deviate from the ideal case.

## 4 Methodology

To comprehensively investigate the impact of various factors on super resolution, a Monte Carlo simulation was conducted. The flowchart depicting the employed methodology is illustrated in Figure 8. This simulation involved an extensive 25,000 iterations, ensuring robustness and reliability in the final results presented in this report. In each iteration, the process initiates with the random sampling of parameters from their respective distributions. These sampled parameters were then inputted into the RS3S/3SA raw image simulator. This simulator effectively generates low-resolution subarray images of the target, incorporating the specified parameters. Subsequently, these low-resolution images are processed through a super resolution algorithm, resulting in the production of a high-resolution image. Following the super resolution step, a resolution measurement was carried out to quantify the effectiveness of the process. This detailed approach allowed for a thorough exploration of the influence of different factors on the super resolution outcome.

Figure 9 presents a comprehensive overview of system parameters and their corresponding values utilized in Monte Carlo simulations. In each iteration, these parameters were sampled from Gaussian distributions, and the resulting resolutions were meticulously documented. This experimental setup allows for a detailed investigation into the influence of individual parameters on super resolution performance. By holding all other factors at their nominal values and selectively varying the parameter under scrutiny, we gain valuable insights into its impact. This systematic approach is crucial in understanding how each parameter contributes to the final high-resolution image, enabling us to fine-tune and optimize the super resolution process for enhanced imaging outcomes. It also provides a robust framework for evaluating the robustness and stability of the system against variations in these parameters.

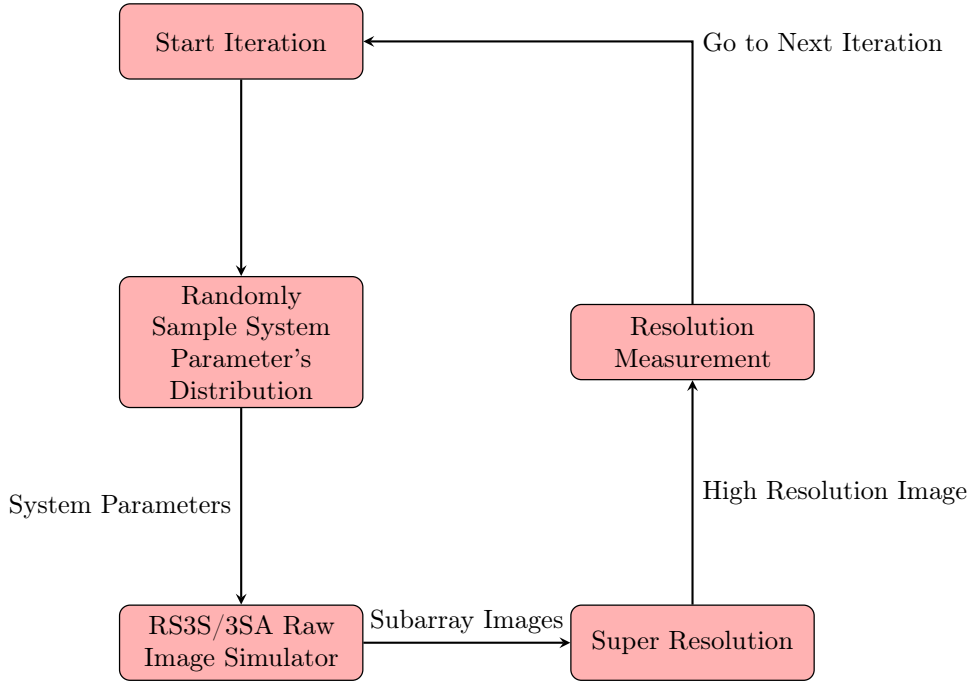

Figure 8: Flowchart of the Methodology Used for Simulations

| Parameter | Nominal Value | Value Range | Sampling |
|---|---|---|---|
| Optics MTF % at 62.5 cycles/mm | 30 | 10-50 | Gaussian |
| Charge Transfer Clock Phase | 1 | 1,2,4 | Uniform |
| Jitter (Pixels) | 0.1 | 0.1-0.2 | Gaussian |
| Signal-to-Noise Ratio at 15% Albedo | 60 | 30-100 | Gaussian |
| Subarray Shifts | 0.5 | 0.1-0.5 | Uniform |
| Error in PSF Estimation (pixels) | Gaussian: $0.5\sigma$ | Gaussian: $1.0\sigma$ Mean: 2 or 3 pixels | Uniform |

Figure 9: Variation of Various Parameters in Monte Carlo Simulations

## 5 Experiments and Results

In the following subsections, we conduct a comprehensive examination of how various imaging system parameters exert influence on the performance of super resolution. This detailed analysis aims to shed light on the intricate interplay between these parameters and the resultant impact on the quality and efficacy of the super resolution process. By dissecting each parameter's contribution, we aim to gain a deeper understanding of their individual and collective effects on achieving high-resolution imaging.

### 5.1 Optical Modulation Transfer Function:

Figure 10 provides a visual representation of how the optics MTF influences super resolution. When the optics MTF is higher, it introduces more aliasing in the low-resolution images, which is a key element for achieving effective super resolution. As the optics MTF value increases, the resulting resolution after super resolution also improves. This enhancement is observed even in scenarios where there may be errors in estimating the high-resolution point spread function, as discussed earlier. This underscores the robustness and effectiveness of the super resolution process in compensating for certain imperfections in the imaging system.

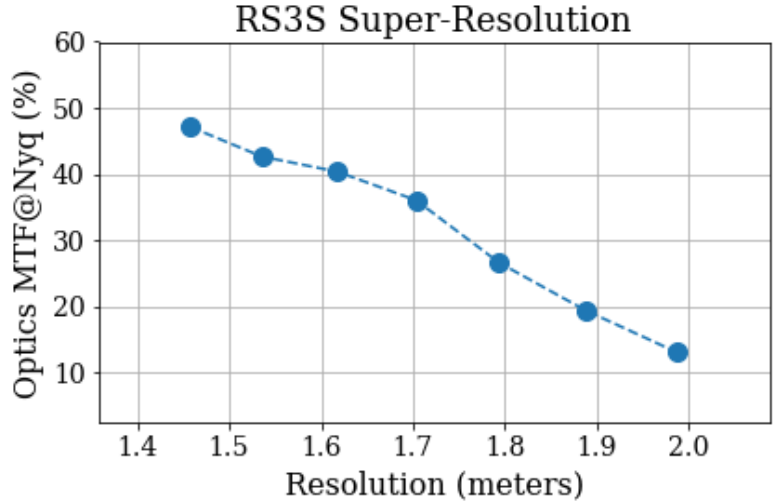

Figure 10: Resolution Sensitivity due to Optics MTF

## 5.2 Signal to Noise Ratio:

For super resolution to be effective, a high SNR is crucial. This high SNR reduces uncertainty and improves the system's ability to detect relevant information by minimizing noise interference. Consequently, as SNR increases, super resolution attains higher resolution, even with potential error in the high-resolution point spread function. Figure 11 provides a visual representation of how the Signal-to-Noise Ratio (SNR) and optical Modulation Transfer Function (MTF) jointly influence super resolution performance. Specifically, optimal performance, depicted in red, is attained when both SNR and MTF values are very high. On the other hand, suboptimal performance, shown in green, is observed when either parameter has a low value. This underscores the equal importance of both factors in achieving effective super resolution.

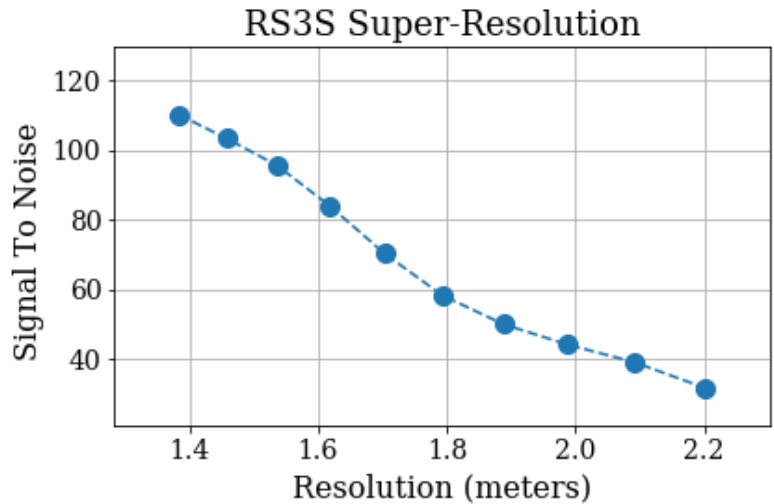

Figure 11: Resolution Sensitivity due to SNR

## 5.3 Optical Modulation Transfer Function & Signal to Noise Ratio:

In the context of the RS3S system, nominal parameters include a 30% optical MTF and a 60 SNR at 15% albedo. These specifications are anticipated to yield an expected resolution of approximately 1.7 meters. To elaborate, the MTF characterizes how well an optical system reproduces fine details in an image. A higher MTF value indicates better performance. The SNR, on the other hand, represents the ratio of the desired signal to background noise. A higher SNR signifies a stronger signal relative to noise interference. In Figure 12, the regions marked in red signify the optimal operating conditions where both SNR and MTF are exceptionally high. In these conditions, the super resolution process can effectively enhance image quality. Conversely, the green regions indicate scenarios where either SNR or MTF, or both, are comparatively low. In such cases, the super resolution process is less effective, resulting in reduced image enhancement. In conclusion, achieving high performance in super resolution necessitates both a high SNR and a robust optical MTF. The RS3S system, with its specified parameters, is expected to yield a resolution of around 1.7 meters under these conditions. This underscores the critical role that both SNR and MTF play in the overall success of super resolution techniques.

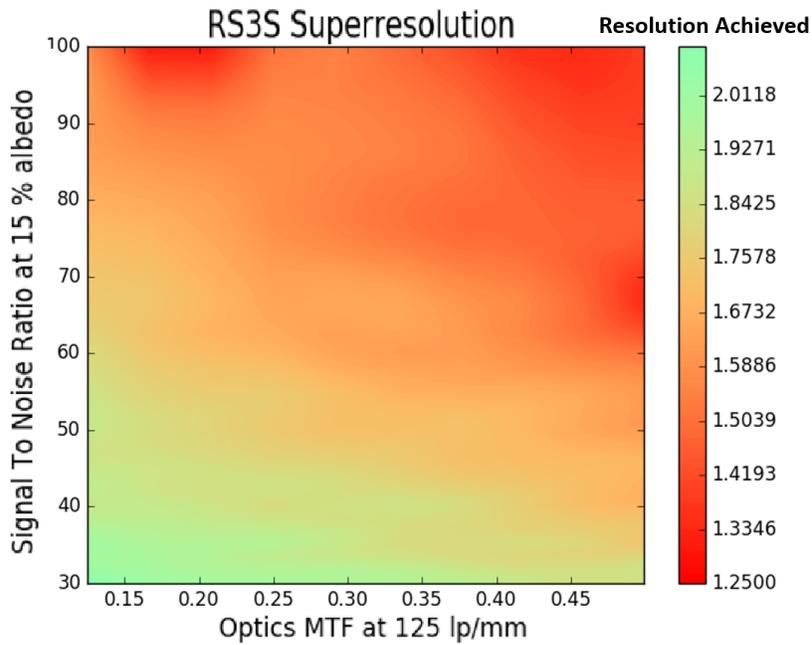

Figure 12: Sensitivity due to MTF and SNR

## 5.4 Jitter:

Figure 13 demonstrates that increasing jitter in the line of sight leads to a decrease in post-super-resolution performance. Line of sight jitter refers to unpredictable shifts in the alignment of optical components or sensors. This can disrupt the super-resolution process, resulting in lower achieved resolution. Minimizing jitter is crucial for optimal super-resolution outcomes.

## 5.5 Clock Phase/ Motion Blur:

Figure14 illustrates how clock phase and motion blur affect low-resolution images and their subsequent super-resolved counterparts. When the number of phases for charge transfer increases, the discrepancy between the charge transfer and satellite velocity diminishes. In simpler terms, this means that the alignment of the charge transfer process and the satellite's movement becomes more accurate. This alignment is crucial for capturing clear images. Moreover, as the clock phases for charge transfer increase, the motion blur decreases. This is significant because reduced motion blur leads to sharper images. When super-resolution is applied to images with less motion blur, it is more effective in achieving higher resolution. In summary, Figure 14 underscores the importance of precise timing in the image capture process. Increasing the number of phases for charge transfer leads to better alignment and less motion blur, ultimately resulting in higher resolution after super-resolution techniques are applied.

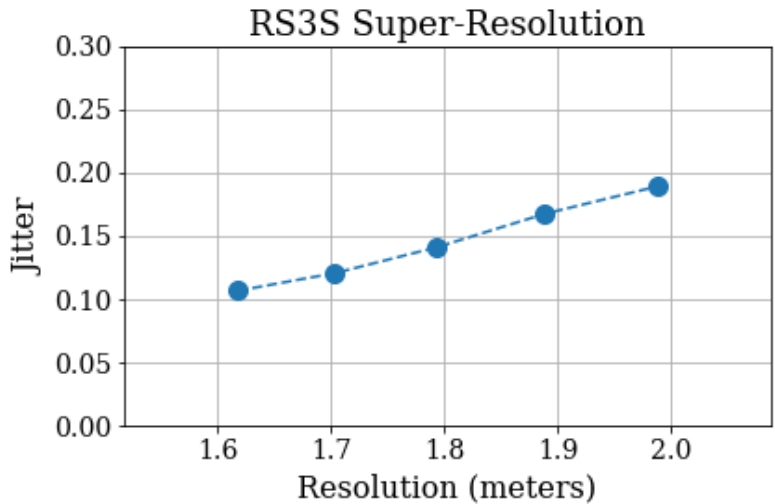

Figure 13: Resolution Sensitivity due to Jitter

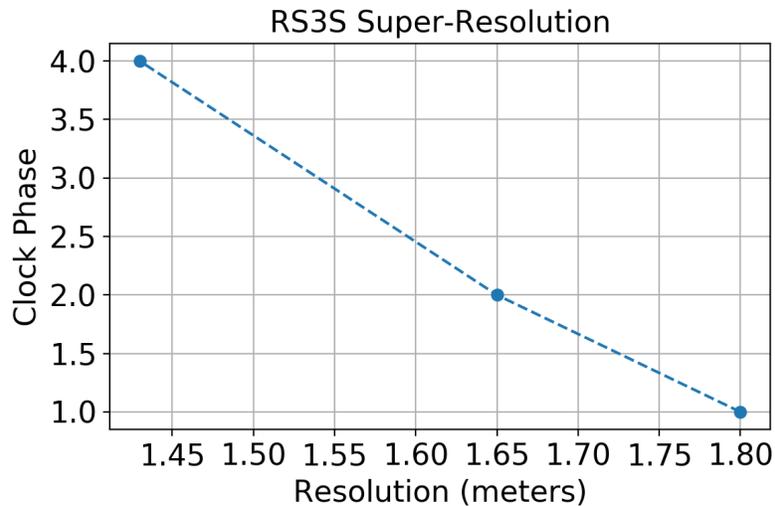

Figure 14: Resolution Sensitivity due to Clock Phases

## 5.6 Subarray Across Track Shifts:

Figure 15 depicts how the super resolution performance changes with alterations in the shifts between two subarrays. It demonstrates that when the subarray shift deviates from 0.5 pixels, the two low-resolution images appear nearly identical due to the absence of a substantial shift between them. Consequently, this leads to a reduction in super resolution performance. To elaborate, consider that subarrays are essentially smaller sections of the overall image. When these subarrays are shifted by a significant amount (in this case, away from the 0.5 pixel mark), they essentially overlap to a large extent. This overlap results in little discernible difference between the two low-resolution images. As a result, the super resolution process has less distinct information to work with, leading to a decrease in its effectiveness. In practical terms, this highlights the importance of having a precise and intentional shift between subarrays for effective super resolution. A shift close to 0.5 pixels provides enough differentiation between the low-resolution images to yield a more significant improvement through super resolution techniques.

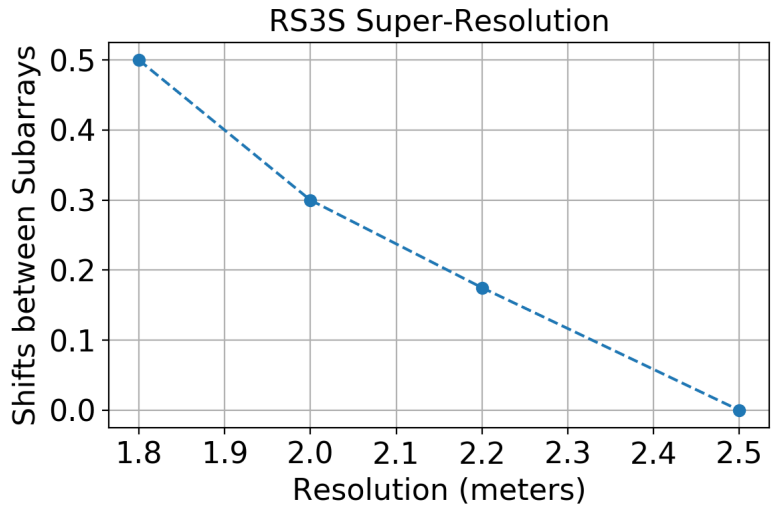

Figure 15: Resolution Sensitivity due to Subarray Shifts

## 5.7 Combined Effect:

Figure 16 displays the histogram representing all the resolutions achieved through the Monte Carlo process. The prominent peak at 1.7 meters indicates that this is the most anticipated resolution achievable with RS3S.

To elaborate, the Monte Carlo process involves running simulations with randomized input parameters to understand the range of possible outcomes. In this case, it's being used to assess the potential resolutions attainable with RS3S. The histogram provides a visual distribution of these outcomes. The peak at 1.7 meters signifies that this particular resolution value occurs most frequently in the simulations. This suggests that, based on the specified parameters and the variability accounted for in the Monte Carlo process, 1.7 meters is the most expected resolution from RS3S.

In practical terms, this information is valuable for setting realistic expectations and understanding the capabilities of the RS3S system under varying conditions. It indicates that, in typical scenarios represented by the Monte Carlo simulations, users can anticipate achieving a resolution around 1.7 meters.

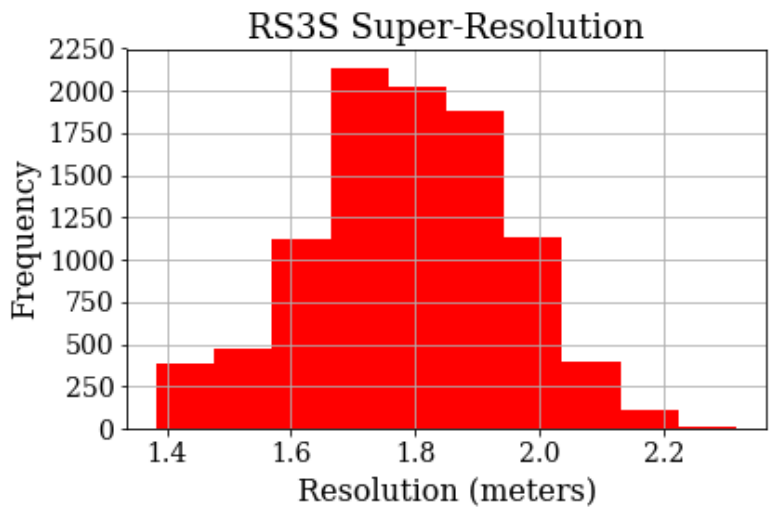

Figure 16: Histogram of Resolutions

## 5.8 Super Resolution Ratio for RS3S/3SA:

Based on the aforementioned study and experiments, it can be deduced that the super resolution factor for AFT images in RS3S is anticipated to be approximately 1.45 times (calculated as 2.5/1.7). This aligns with the standard provided products in missions globally, including SkySat, which offers a super resolution product with a factor of approximately 1.44 times (derived from a native resolution of 0.72m to 0.5m).

Applying this factor of 1.45 times to super resolution, the projected resolution for FORE images in RS3S is anticipated to be around 2.20m (AL) x 1.90m (AX), considering the Ground Sampling Distance (GSD) of 3.2m (AL) x 2.8m (AX). Given that the primary product from RS3S/3SA is a digital elevation model, with a targeted horizontal spacing of 5m and a vertical accuracy of 5m, this level of resolution is deemed sufficient to meet the specified requirements.

# 6 Conclusion

Super resolution proves to be a pivotal technique in augmenting the resolution capabilities of the RS3S/3Sa. Through a comprehensive analysis, we have brought crucial insights into the interplay of various parameters. With an optimal SNR and MTF, the system is poised to achieve its highest potential resolution. In particular, a 60 SNR coupled with a 30% optical MTF yields an expected resolution of approximately 1.7 meters. Moreover, minimizing line of sight jitter emerges as a critical consideration. The system's stability is paramount, as significant jitter can significantly hinder the effectiveness of super resolution techniques. Precise subarray shifts also play a pivotal role. Appropriate shifts ensure that the super resolution process can effectively leverage available information, resulting in a notable enhancement in resolution.

The findings from the study and experiments suggest that RS3S is projected to achieve a super resolution factor for AFT images in line with globally provided mission products, including SkySat. This factor, approximately 1.45 times, demonstrates promising potential for enhanced image quality. Additionally, the anticipated resolution for FORE images, at 2.20m (AL) x 1.90m (AX), aligns well with the specified requirements for digital elevation modeling.

In conclusion, the RS3S system, with its calibrated parameters and meticulous attention to stability and alignment, demonstrates its capability to achieve high-resolution imagery. Under typical conditions, users can expect to achieve a resolution of around 1.7 meters through the judicious application of super resolution techniques. This enhancement significantly bolsters the system's remote sensing capabilities, providing a powerful tool for various applications in Earth observation and beyond.